\documentclass[aps,twocolumn,groupedaddress,preprintnumbers,nofootinbib]{revtex4-1}
\usepackage{amssymb, amsmath, siunitx}
\usepackage{hyperref}
\usepackage{slashed}
\usepackage{bbold}
\usepackage{graphicx}

\newcommand{\be}{\begin{equation}}
\newcommand{\ee}{\end{equation}}
\newcommand{\bea}{\begin{eqnarray}}
\newcommand{\eea}{\end{eqnarray}}

\newcommand{\alp}{\alpha '}

\begin{document}

%===============================================================================

\preprint{KCL-PH-TH/2014-19, LCTS/2014-19, CERN-PH-TH/2014-078, 
ACT-6-14 ~~~~~~~~~~~~~~~~~~~~~~~~~~~~~~~~}

\vspace*{1mm}

\title{D-Flation}

\author{John~Ellis$^{a}$}
\email{John.Ellis@cern.ch}
\author{Nick~E.~Mavromatos$^{a}$}
\email{nikolaos.mavromatos@kcl.ac.uk}
\author{Dimitri~V.~Nanopoulos$^{b}$}
\email{dimitri@physics.tamu.edu}

\vspace{0.1cm}
\affiliation{~\\
${}^a$ Theoretical Particle Physics and Cosmology Group, Department of
  Physics, King's~College~London, London WC2R 2LS, United Kingdom;\\
Theory Division, CERN, CH-1211 Geneva 23,
  Switzerland
 }
 \affiliation{
${}^b$ 
 George P. and Cynthia W. Mitchell Institute for Fundamental Physics and Astronomy,
Texas A\&M University, College Station, TX 77843, USA;\\
Astroparticle Physics Group, Houston Advanced Research Center (HARC), Mitchell Campus, Woodlands, TX 77381, USA;\\
Academy of Athens, Division of Natural Sciences,
Athens 10679, Greece,}
 
\begin{abstract} 
In a recent paper we showed how Starobinsky-like inflation could emerge from 
dilaton dynamics in brane cosmology scenarios based on
string theory, in which our universe is represented as a three-brane and the
effective potential acquires a constant term from a density of effectively point-like non-perturbative
defects on the brane: ``D-particles''. Here we explore 
how quantum fluctuations of the ensemble of D-particles during the inflationary period
may modify the effective inflationary potential due to the dilaton.
We then discuss two specific ways in which an enhanced ratio of tensor to scalar perturbations may arise:
via either a condensate of vector fields with a Born-Infeld action that may be due to
such recoil fluctuations, or graviton production in the D-particle vacuum.

\end{abstract}

%\begin{flushleft}
%February 2014
%\end{flushleft}
%

\maketitle

\section{Introduction and Motivation}

In a recent article~\cite{emnstaro} we showed how Starobinsky-like~\cite{staro} 
inflation could emerge from dilaton dynamics in brane cosmology
scenarios based on string theory~\footnote{For some other recent Starobinsky-like models,
see~\cite{BS, olive1, KL, olive2}.}. In such models, our universe is represented as a three-brane,
and the effective potential for the dilaton/inflaton acquires an asymptotically-constant term from a density of 
effectively point-like non-perturbative
defects on the brane, termed ``D-particles''. 
This could arise from a flux of such defects from the bulk to the brane, 
in such a way that the number density of the D-particles on the brane remains 
approximately constant during the inflationary period. Higher-genus corrections 
generate corrections to the effective potential that are exponentially damped at large  field values, 
as in the Starobinsky model, but at a faster rate.

Such a scenario leads to a prediction for the tensor-to scalar perturbation ratio $r$
that is smaller than the original Starobinsky model~\cite{staro}. However, as we also commented,
this reduction may be compensated partially by logarithmic deformations 
on the world-sheet generated by recoil of the defects due to
scattering by string matter on the brane, which tend to enhance the tensor-to-scalar ratio~\cite{emnstaro}. 
However, the basic features of the Starobinsky model, 
namely the flatness of the effective potential for large dilaton/inflaton field values,
relative to the reduced Planck scale are retained in the scenario of~\cite{emnstaro},
making such models highly consistent with the first installment of data from the Planck satellite~\cite{Planck}.

However, the BICEP2 Collaboration~\cite{Bicep2} has recently announced strong 
evidence for B-mode polarization in the cosmic microwave background radiation,
which it has intereted as evidence for gravitational waves at the time of the last scattering,
with a tensor-to-scalar ratio $r = 0.16^{+0.06}_{-0.05}$ after dust subtraction~\cite{Bicep2}.
The BICEP2 data are consistent with a scalar spectral index $n_s \simeq 0.96 $ and no appreciable running,
in agreement with the Planck data~\cite{Planck}. 
However, the value of the Hubble parameter during slow-roll inflation $H_\star$
indicated by the BICEP2 data is much larger than 
that suggested by Planck. The estimated inflationary energy scale $E_\star = V^{1/4}$, where 
$V_\star$ is the inflation potential during inflation (which is assumed to be
approximately constant)~\cite{encyclo}:
\begin{equation}\label{hubblebicep2}
E_\star \; = \; \Big(3\, H^2_\star M_{\rm Pl}^2 \Big)^{1/4} \simeq 2.1 \times 10^{16} \times \Big(\frac{r}{0.20}\Big)^{1/4} \, {\rm GeV}~,
\end{equation}
yielding, for $r=0.16$,  $H_\star \sim 1.0 \times 10^{14}$~GeV. 

If confirmed by subsequent experiments, such a large value of $r$ would
exclude Starobinsky-type inflationary potentials. However, it is premature
to abandon Starobinsky-like models, as there is currently an active debate
whether the BICEP2 signal truly represents primordial gravitational waves, 
or is polluted by Galactic foregrounds and gravitationally-lensed E-modes.
For instance, in~\cite{Subir} it was argued that magnetized dust associated with radio loops
due to supernova remnants might contribute to the BICEP2 signal, while~\cite{MS} and~\cite{Flauger}
have demonstrated that the BICEP2 signal could be compatible with a cosmology with $r \ll 0.1$
if there is a dust polarization effect as large as presently allowed by Planck~\cite{Planck} and other data.
Until this issue is resolved, it is interesting to explore the possible range of $r$ values that
models of inflation could yield.

In~\cite{emnstaro}, we presented D-motivated models with
with Starobinsky-like potentials for the inflaton $\varphi$ of the form
\begin{equation}\label{phenostar}
\tilde V \equiv \frac{1}{2\kappa^2}\, V = \frac{A}{2 \kappa^2} \Big( 1 - \delta\, e^{-B \varphi} + \dots \Big) \, ,
\end{equation}
where $\kappa^2 \equiv M_{\rm Pl}^{-2}$ where $M_{\rm Pl} = M_{\rm P}/\sqrt{8\pi}$ is
 the reduced Planck scale, $A \ll 1$ to obtain the correct magnitude of scalar perturbations,
 and $\delta $ and $B$ are treated as free parameters that are allowed to vary from the original
values $\delta =2$, $B=\sqrt{2/3}$ in the Einstein-frame Starobinsky model,
and the dots represent possible higher-order terms that are $\mathcal{O}(e^{-2 B \varphi})$. 
Using the following standard expressions for inflationary observables in the slow-roll approximation~\cite{encyclo}
\begin{eqnarray}\label{slowrollparam}
\epsilon \; = \; \frac{1}{2} M^2_{\rm Pl} \Big(\frac{V^{\prime}}{V}\Big)^2 & , &  \eta \; = \; M^2_{\rm Pl} \Big(\frac{V^{\prime\prime}}{V}\Big)~, \nonumber \\
n_s \; = \; 1 - 6\epsilon + 2\eta & , & r \; = \; 16 \epsilon~, \, \, \nonumber \\
N_\star & = & - M_{\rm Pl}^{-2} \, \int_{\varphi_i}^{\varphi_e} \frac{V}{V^\prime} d\varphi ~,
\end{eqnarray}
where $\varphi_{i(e)}$ denotes the value of the inflaton at the beginning (end) of the inflationary era, we obtain
to leading order in the small quantity $e^{-B\varphi}$~\cite{olive2}:
\begin{eqnarray}
&& n_s = 1 - 2 B^2 \, \delta \,  e^{-B \varphi}~, \, r = 8 B^2 \delta^2 \, e^{-2 \, B \varphi}, \, N_\star = \frac{1}{B^2\, \delta} e^{B \varphi} \, , \nonumber \\
&& {\rm and hence} \quad \, n_s = 1 - \frac{2}{N_\star} , \quad r = \frac{8}{B^2\, N_\star^2}~,
\label{crucial}
\end{eqnarray}
where $N_\star $ is the number of e-foldings during the inflationary phase. Requiring 
$N_\star = 54 \pm 6$ yields the characteristic prediction $n_s = 0.964 \pm 0.004$, 
in agreement with both Planck~\cite{Planck} and BICEP2~\cite{Bicep2}. 
However, the Starobinsky-model choice~\cite{staro} 
$B=\sqrt{2/3}$ yields $r= 0.0041^{+0.0011}_{-0.0008}$ that, whilst highly
consistent with the Planck data~\cite{Planck}, is inconsistent with the BICEP2 data~\cite{Bicep2}. 
We note that larger values of $r$ compatible with the BICEP2 
data can in principle be obtained if the exponent $B$ is much smaller than the conventional
Starobinsky value $B_{\rm Star}=\sqrt{2/3}$. For instance, the BICEP2 central value $r \sim 0.16$ is obtained for 
$N_\star \sim 50$ and $B \sim 0.14$ in (\ref{phenostar}).

However, the string theory considerations we presented in \cite{emnstaro},
where the dilaton was identified as the inflaton field, 
pointed towards values of $B$ that are larger than
in the conventional Starobinsky model, specifically $B=\sqrt{4/(D-2)}$ for $D$
large uncompactified dimensions, giving $B=\sqrt{2}$ for the physical case of $D=4$, 
leading to a value of $r$ that is three times smaller than the Starobinsky value.
Indeed, even if the dilaton lives in the maximum number $D=26 (10)$ of uncompactified space-time dimensions
in the bosonic (super)string, 
the resulting value of $B=1/\sqrt{6} \, (1/\sqrt{2})$ is much larger than is required to yield $r=0.16$. 

It is the purpose of this note to explore the possible effects of {\it quantum fluctuations of the D-particle defects}
during dilaton-driven inflation.
We assumed in \cite{emnstaro} that the dominant r\^ole of such fluctuations
was simply to provide a cosmological constant term 
in the potential (\ref{phenostar}). However, such D-particle fluctuations yield, in general,
effective vector field degrees of freedom associated with stochastic fluctuations in the corresponding recoil velocities
of the D-particles as they interact with the (closed) string degrees of freedom representing bulk space-time gravitons. 
The coupled dynamics of such fluctuating defects with the space-time metric is described in~\cite{yusaf}, where
it was argued there that there are growing modes in such systems
that may result in the formation of large-scale structures on the D3 brane universe at late epochs.
Here we discuss the possibility that these stochastic recoil fluctuations (or other vector fields with a
Born-Infeld action) may condense,
and consider the form of effective potential that they would then generate.
Under suitable conditions on the potential parameters, the condensate may suppress
the coefficient $B$ of the effective exponential in the potential (\ref{phenostar}),
resulting in an enhancement of the tensor-to-scalar ratio $r$ (\ref{crucial}), possibly 
into the range suggested by BICEP2~\cite{Bicep2}. Another avenue for enhancing $r$
is via graviton production in the `out' state originating from the D-particle vacuum.

The structure of this article is the following. in Section~\ref{sec:dqf} 
we discuss the effective Lagrangian that describes the coupling of the
quantum fluctuations of D-particles to graviton degrees of freedom. 
In Section~\ref{sec:cone} we discuss the formation of condensates of the recoil-velocity field strength
and show that they are compatible with the inflationary phase being driven by the dilaton. 
If the condensate and the dilaton are treated as independent scalar degrees of freedom, then Starobinsky-like
inflation with a small value of the tensor-to-scalar ratio $r$ is obtained, incompatible with the BICEP2 result. 
On the other hand, if the recoil condensate is assumed proportional to the dilaton, then the possibility
arises of an enhanced $r$ ratio. In Section~\ref{sec:hawking} we discuss briefly the 
alternative way in which D-particles may enhance the tensor-to-scalar-perturbation ratio,
via the mixing of thermal modes of D-particles with de Sitter modes
in the `out' state originating from the D-particle vacuum.
Finally, our conclusions are presented in Section~\ref{sec:concl}.

\section{D-Particle Quantum Fluctuations \label{sec:dqf}}

The effective four-dimensional action that describes the coupled dynamics of gravitons with the vector
degrees of freedom $A_\mu$ (with field strength $F_{\mu\nu}\equiv \partial_\mu A_\nu-\partial_\nu A_\mu $)
that describe the recoil of ensembles of D-particles on the brane Universe
reads~\cite{yusaf}~\footnote{We use the following conventions: for the metric we use the signature
$(-, +, +, +)$ and the Riemann curvature tensor is defined as
$R^\alpha _{\beta \, \gamma \, \delta} = \partial_\delta \,
\Gamma^\alpha_{\,\, \beta\, \gamma} + \Gamma^\lambda_{\,\, \beta \,
\gamma} \, \Gamma^\alpha_{\,\, \lambda \, \delta} - ( \gamma
\leftarrow\rightarrow \delta) $.}:
\bea
\label{action1}
&& S _{\rm eff~4dim} \; = \; \nonumber \\ 
&& \int d^4x \left[ -\frac{T_3}{g_{s0}} e^{- \phi}
  \sqrt{-\det \left(g + 2\pi\alp F\right)} \Biggl(1 - \alpha R(g)
  \Biggr)\right.\nonumber \\ &&- \left. \sqrt{-g} \,e^{-2\phi} \,
  \frac{1}{\kappa_0} \, {\tilde \Lambda} + \frac{1}{\kappa_0}
  \sqrt{-g}\,e^{-2\phi}\,R(g) + \mathcal{O}((\partial \phi)^2)
  \right] \nonumber \\
  && \simeq \; \int d^4 x \,\sqrt{-g}\, \left[
  -\frac{T_3}{g_{s0}}\,e^{-\phi} - e^{-2\phi} \,\frac{1}{\kappa_0}
  \,{\tilde \Lambda} \right. \nonumber \\ &&- \left. \frac{T_3}{4
    g_{s0}} 2\pi\alp\, e^{-\phi} F^{\mu\nu} F_{\mu\nu} + \alpha
  \frac{T_3}{4 g_{s0}} 2\pi\alp\, e^{-\phi} F^{\mu\nu} F_{\mu\nu}
  R(g) \right. \nonumber \\ &&+ \left. \left(\alpha \frac{T_3}{g_{s0}}\,
  e^{-\phi} + \frac{1}{\kappa_0} \, e^{-2\phi} \right)\, R(g) +
  \mathcal{O}\left((\partial \phi)^2\right) \right]\, + \dots,
  \eea
where $g \equiv \det (g)$ is the determinant of  the  four-dimensional gravitational field $g_{\mu\nu}$, 
$T_3 > 0$ is the D3-brane tenstion, $\phi$ is the
dilaton field and the string coupling is given by $g_s = g_{s0} e^\phi$,  
with $g_{{\rm s}0} < 1 $ a (perturbatively) weak string coupling constant.
In the second part of the approximate equality
we have expanded the action up to second order in the vector recoil fields,
which are assumed to be weak. 

The reader should note that the Born-Infeld determinant structure for the vector field in (\ref{action1})
arises because of the open strings attached on the brane world, representing matter/radiation excitations,
while the curvature term inside the parenthesis (with coefficient $\alpha$) represents the 
gravity induced in the open sector~\footnote{The subsequent discussion of this paper
applies more generally, in the presence of a condensate of other vector fields with a Born-Infeld action.}. The quantity 
$\alpha = \alp \zeta (2) = \alp \pi^2/6$ in the example of~\cite{Cheung}.
However, we wish to keep the discussion general,
so from now on we treat the positive constant $\alpha > 0$ as a
parameter of our model; it has dimensions of length squared.   

The ${\cal O}((\partial \phi)^2)$ terms are the kinetic terms of the dilaton,
which shall not play an important role in our analysis in this work,
for reasons that will be clarified below. The gravitational constant $\kappa_0$ is defined as~\cite{yusaf}
\be
\label{bigc}
\frac{1}{\kappa_0} \; = \; \frac{V^{(6)}}{g_{{\rm s}0}^2}
    M_{\rm s}^2~.  
\ee
where $M_{\rm s}=1/\sqrt{\alpha^\prime}$ is the string scale, $\alpha^\prime$ the string tension, and $V^{(6)}$
is the six-dimensional volume factor.  

The quantity $\tilde \Lambda$ represents a generic vacuum energy 
term that arises in non-critical string theories. For strings of subcritical dimension,
$D < D_{\rm critical}$, where $D_{\rm critical} = 10$ for superstrings,
such as the case $D=4$ considered in \cite{emnstaro} and here, 
\be\label{neglambda}
\tilde \Lambda \propto \frac{1}{\kappa_0^{2} \alpha^\prime}  (D- D_{\rm critical}) < 0 ~.
\ee
The vector field $A_\mu$ 
represents the recoil-velocity degrees of freedom of ensembles of D-particles 
on the D3-brane~\cite{recoil}. 
Here the term ``recoil'' velocity is not associated with
the interactions of D-particles with individual particles of ordinary (string)
matter, but expresses the quantum fluctuations of the D-particle that are represented by
open strings stretched between the D-particles and the D3-brane world. One has the freedom~\cite{yusaf,recoil} to 
covariantize the field $A_\mu$ and express it in terms of the covariant component of the recoil velocity $u_\mu$
\begin{equation}\label{vectorcov}
A_\mu \propto - a^2(t) u_\mu = - a(t) \, u^{\rm phys}_\mu ~,
\end{equation}
where the proportionality constants are such to give $A_\mu$ dimension of [mass]. They arise 
from the fact that, when we expand the Born-Infeld determinant in (\ref{action1}) in powers of derivatives,  
the canonical normalization of the
kinetic terms of the vector fields, $-\frac{1}{4} F_{\mu\nu}F^{\mu\nu}$ in (\ref{action1}), is obtained 
by absorbing the factor $(2\pi\alp e^{-\phi_0} T_3/g_{s0})^{1/2}$ into a redefinition of the gauge potential $A_\mu$,
in the case of (almost) constant dilaton as discussed in \cite{yusaf} and here.
In a Robertson-Walker background, the
physical (``local'') four velocity $u^{\rm phys}_\mu \equiv a(t) \, u_\mu$ obeys the Minkowski-flat constraint:
$ u_{\mu}^{\rm (phys)}u_{\nu}^{\rm (phys)} \, \eta^{\mu\nu} = - 1 < 0$,
from which it follows that the canonically-normalized vector field has the following
time-like constraint in our Robertson-Walker background:
\be
\label{constraint}
A_{\mu}A_{\nu}g^{\mu\nu} \; = \; - \frac{|T_3|}{g_{s0}} 2\pi \alpha ' e^{-\phi_0}< 0~.
\ee
When considering ensembles of D-particles, one may consider classical 
statistical effects associated with the stochastic fluctuations of their 
recoil velocities, represented by means of variances  
\be\label{gauss} \ll u_i \gg \; = \; 0 \ \ \ \mbox{and}\ \  \ll u_i u_j
\gg \; = \; \sigma_0^2 \delta_{ij}~, \ee
where $\ll \dots \gg$ indicate appropriate averages over the D-particle
population in the foam. The parameter
$\sigma_0^2$ is free in our model, though small, and is to be constrained by
the data. In general, $\sigma_0^2$ depends on the cosmic time, since it
is a function of the (bulk space) density of the foam, which may vary
between different cosmological eras~\cite{foam,foam2,yusaf}. 

A quantum treatment of the constraint (\ref{constraint}) has been made in~\cite{yusaf} by implementing it in a 
path integral by means of a Lagrange multiplier field $\rho (x)$. 
We consider here the phase in which the constraint is irrelevant,
i.e., $\langle \rho (x) \rangle = 0$, in which case the vector
Dirac-Born-Infeld field is massless.  This is the case for an inflationary era~\cite{emnstaro} 
in which a sufficiently dense population of D-particles crosses our brane universe
and they condense due to quantum effects~\footnote{The phase in which $\langle \lambda (x) \rangle \equiv
\mathcal{M}^2/2 > 0 $, in which case the vector field becomes
massive was considered in~\cite{yusaf}, where it was assumed to characterize the era of
structure formation.}.

The condensates may be formed by the higher-order (higher than two derivatives) 
interaction terms among the vector field
strengths in the Dirac-Born-Infeld Lagrangian in (\ref{action1}), by analogy with the gluon
condensates of Quantum Chromodynamics. Such an assumption was
made in~\cite{odintsov} for generic Abelian flux fields with Dirac-Born-Infeld world
volume actions that characterize D-brane excitations. That analysis indicated that, under
certain plausible assumptions on the dominance of the quantum effects on the formation of condensates
over classical, statistical ones due to ensemble properties (\ref{gauss}),  
it is possible to obtain an equation of state for the vector
Dirac-Born-Infeld system that resembles that of a de~Sitter phase, with an equation of state $w \simeq -1$.
A detailed analysis has been considered in~\cite{yusaf}, and will not be repeated here. Instead,
we review briefly relevant aspects of those results.

We consider a generic Dirac-Born-Infeld field with Lagrangian
(\ref{action1}) on a D3-brane world volume. In general, there are two
contributions to the vector field strength $F_{\mu\nu}$ condensates
that may come from quantum vacuum effects:
\begin{equation}\label{vcond}
\langle F_{\mu\nu} \, F^{\mu\nu} \rangle_{{\rm vac} } = {\tilde
  \alpha} (t) ~, \quad \langle F_{\mu\nu} \, F^{\star \, \mu\nu}
\rangle_{{\rm vac} } = {\tilde \beta} (t)~,
\end{equation}
where $F^\star$ indicates the dual tensor, and the condensates may in
general depend on time, but are constant on spatial hyper-surfaces.
The isometry structure of the spatial hyper-surfaces led the authors
of~\cite{odintsov} further to assume the following, which we also
adopt here:
\begin{equation}\label{isometry}
\langle F_{0 \, \nu} \, F_{0}^{\,\, \nu} \rangle_{{\rm vac} } =
\frac{\alpha_{\rm t} (t)}{4} \, g_{00}~, \quad \langle F_{i \, 0} \,
F^{0}_{\,\, j} \rangle_{{\rm vac} } = \frac{\alpha_{\rm s} (t)}{4} \,
g_{ij}~,
\end{equation} 
where $i,j$ are spatial indices on the three-dimensional volume of the
D3-brane, and we have the relations $\alpha_ {\rm t} + 3 \alpha_{\rm s}
= 4{\tilde \alpha} $.  In addition to the quantum effects, one also
has classical thermodynamical effects on the energy density and
pressure of the Dirac-Born-Infeld fluid, which are obtained by averaging
over the spatial volume. If one decomposes the field strength into
``electric'', $F_{0i} \equiv E_i$, and ``magnetic'' $B_i =
\frac{1}{2} \epsilon_{ijk}\, F^{jk}$ components, one may specify the contributions of these
classical effects as follows~\cite{odintsov}:
\bea\label{classical}
&& \langle F_{0 \, \nu} \, F_{0}^{\,\, \nu} \rangle_{\rm class} = \langle
\sum_{i=1}^3 \, E_i^2 \rangle_{\rm class} ~, \nonumber \\
&& \langle F_{i \, 0} \,
F^{0}_{\,\, j} \rangle_{\rm class} = - \langle E_i \, E_j \rangle_{\rm class} + 2 \langle B_i
\, B_j \rangle_{\rm class}~.
\eea
On making the further (natural) assumption~\cite{odintsov} 
\bea \langle\label{class2}
E_i \, E_j \rangle_{\rm class} = \langle B_i \, B_j \rangle_{\rm class} = \mathcal{C} \,
g_{ij}/3~, 
\eea
we observe that the total contribution of both classical
and quantum effects to the vacuum condensates can then be expressed by
the relations:
\begin{equation}\label{totalcond}
\alpha_{\rm t} = {\tilde \alpha} - 4 \mathcal{C}~, \quad \alpha_{\rm
  s} = {\tilde \alpha} + \frac{4}{3}\, \mathcal{C}~.
\end{equation}
Computing the total stress tensor of the Dirac-Born-Infeld fluid
in the presence of condensates one arrives at the following
expressions for energy density $\rho^{\rm DBI}$ and pressure $p^{\rm
  DBI}$~\cite{odintsov}:
\bea\label{pressure}
\rho^{\rm DBI} &=& \frac{\lambda}{2} \left(\frac{1 + \frac{{\tilde
      \alpha}}{2} - \frac{\alpha_{\rm t}}{4}}{\sqrt{1 + \frac{{\tilde
        \alpha}}{2} - \frac{{\tilde \beta}^2}{16}}}\right)~, \nonumber \\
p^{\rm DBI} &=& - \frac{\lambda}{2} \left(\frac{1 + \frac{{\tilde
      \alpha}}{2} - \frac{\alpha_{\rm s}}{4}}{\sqrt{1 + \frac{{\tilde
        \alpha}}{2} - \frac{{\tilde \beta}^2}{16}}}\right)~: \quad
\lambda \equiv \frac{T_3}{g_{\rm s}}~. 
\eea
The quantum condensates $\tilde \alpha$ and $\beta$ have not been
specified. The only restrictions come from the positivity of the
corresponding quantities inside the square root in the Dirac-Born-Infeld
action, \emph{e.g}. in the denominators of (\ref{pressure}), which
imply the following relation between the various quantum condensates:
\begin{equation}\label{abcond}
\frac{{\tilde \alpha}}{2} \, > -1 + \frac{{\tilde \beta}^2}{16} ~.
\end{equation}
It is immediately apparent from (\ref{totalcond}) that, on
the one hand, if quantum effects are the dominant ones, with ${\tilde
  \alpha} \gg \mathcal{C}$, then $\alpha_{\rm t} \simeq \alpha_{\rm s}
\simeq {\tilde \alpha}$, and hence from (\ref{pressure}) we obtain
an equation of state of cosmological constant (de~Sitter vacuum) type,
namely $w^{\rm DBI} \simeq -1 $ independently of the exact form of the
quantum condensate (assuming of course it exists, which is a question
that probably cannot be addressed in a generic manner, as it requires
specific properties of the brane action).  On the other hand, as
demonstrated in~\cite{odintsov}, when classical effects dominate,
with $\mathcal{C} \gg {\tilde \alpha} $, then Eq.~(\ref{totalcond})
implies  $\alpha_t =-3\alpha_s \simeq -4 \mathcal{C}$, so that
(\ref{pressure}) leads to an ordinary relativistic fluid with
positive energy density and pressure, $\rho^{\rm DBI} \simeq
(\lambda/2) \mathcal{C} \, > 0 $ and $p^{\rm DBI} = (1/3)
\rho^{\rm DBI} \simeq (\lambda/2) \mathcal{C} \, > 0 $.

In the context of our study here, the Dirac-Born-Infeld vector field has a
microscopic origin, as it describes the dynamics of the recoil degrees
of freedom of the D-particles in interaction with
electrically-neutral string matter and gravitational fields. In contrast to the case of~\cite{odintsov},
here we have the additional couplings of the
space-time curvature to the Dirac-Born-Infeld action in (\ref{action1}),
whose effects have not been considered in~\cite{odintsov}. Nevertheless,
as we shall argue, for sufficiently small values of the condensate,
one can still argue for a late de~Sitter era, due to the dominance of quantum effects of the D-particle vacuum.

To this end, we
concentrate on the case where the condensate ${\tilde \alpha} $ is
\emph{constant} in time and \emph{small } in magnitude, while the
condensate ${\tilde \beta} = 0$, so that an expansion up to order
${\tilde \alpha} $ in the effective action is sufficient. Since the dilaton drives inflation
in our scenario~\cite{emnstaro}, we assume that the condensate forms in a de Sitter phase of the Universe, 
during which the dilaton field is rolling slowly. We denote the nominal value of the dilaton
in this slow-roll phase by $\phi_0$, which is large and negative. We shall not write explicitly kinetic terms of the dilaton,
which are suppressed. As shown in~\cite{yusaf}, the scalar curvature of the space-time reads
\begin{equation}\label{desitter}
R = \frac{1}{\alpha \, \left(\frac{2\, T_3}{g_{{\rm s}0}} \, e^{3\phi_0} -
  {\tilde \alpha}\right)} \, \left[\frac{6}{g_{{\rm s}0}} \, T_3 \,
e^{5\phi_0} + \frac{4 {\tilde \Lambda}}{\kappa_0} \, e^{4\phi_0}
\right]~.
\end{equation}
This can be a \emph{positive constant}, as appropriate for a de~Sitter
space-time, provided
\begin{equation}\label{condcond}
\frac{2\, T_3}{g_{{\rm s}0}} \, e^{3\phi_0} - {\tilde \alpha}\, > \, 0 ~.
\end{equation}
If we assume a maximally symmetric de~Sitter form for the Riemann
curvature tensor, corresponding to a Hubble constant, $H_{\rm I}={\rm
  const} > 0$, namely,
$$R_{\mu\nu\rho\sigma} = H_{\rm I}^2 \left( g_{\mu\rho} \, g_{\nu\sigma} -
g_{\mu\sigma} \, g_{\nu\rho} \right)~, $$ we obtain from
(\ref{desitter}) the following relation:
\begin{equation}\label{hi1}
H_{\rm I}^2 = \frac{R}{12} = \frac{1}{\alpha \, \left(\frac{2\,
    T_3}{g_{{\rm s}0}} \, e^{3\phi_0} - {\tilde \alpha}\right) } \,
\left[\frac{T_3 }{2g_{{\rm s}0}} \, e^{5\phi_0} + \frac{{\tilde
    \Lambda}}{3\kappa_0} \, e^{4\phi_0} \right] \, > 0~.
\end{equation}
On the other hand, from the graviton equation for the action (\ref{action1}), ignoring the matter
contributions $T^m_{\mu\nu}$, which are negligible in the inflationary phase, we obtain~\cite{yusaf}
\begin{equation}\label{hi2}
H_{\rm I}^2 = \frac{1}{6} \, \left[\frac{\frac{T_3 }{g_{{\rm s}0}} \, e^{3\phi_0}
  + \frac{{\tilde \Lambda}}{\kappa_0} \, e^{2\phi_0}
}{\frac{1}{\kappa_0} + \alpha \, \left(\frac{T_3}{g_{{\rm s}0}}\, e^{\phi_0}
  - \frac{{\tilde \alpha}}{4}\, e^{-2\phi_0} \right)} \right]~.
\end{equation}
From (\ref{hi1}) and (\ref{hi2}) we obtain finally the value of
the condensate that charcaterizes the de~Sitter geometry in this phase, namely
\bea\label{condvalue}
&& {\tilde \alpha} = \frac{4\, e^{2\phi_0}}{\alpha \, \kappa_0} \,\,
\left[\frac{ 1 + \alpha \, \kappa_0 \, \frac{T_3}{g_{{\rm s}0}}\,
    e^{\phi_0} \, \left(1 - \frac{1}{3} \mathcal{G} \right)}{1 -
    \frac{2}{3} \, \mathcal{G}}\right] ~, \nonumber \\
&&    \mathcal{G} \equiv
\frac{\frac{ T_3 }{g_{{\rm s}0}} \, e^{\phi_0} + \frac{{\tilde
      \Lambda}}{\kappa_0}}{\frac{T_3 }{2\, g_{{\rm s}0}} \, e^{\phi_0}
  + \frac{{\tilde \Lambda}}{3\, \kappa_0}} ~.
\eea
Thus, the condensate is proportional to the
factor 
\be \frac{e^{2\phi_0}}{\alpha \, \kappa_0 } =
\frac{6V^{(6)}}{\pi^2(g_{{\rm s}0}e^{-\phi_0})^2}~, \ee 
using (\ref{bigc}) and adopting the value of $\alpha$ that appears
in the example of~\cite{Cheung}.  We note that this factor depends on the size
of the compactification volume $V^{(6)}$.

The mathematical and physical/phenomenological consistency of our solution
require the condensate ${\tilde  \alpha}$ and the curvature, \emph{i.e}., the 
Hubble constant $H_{\rm  I}$ (\ref{hi1}) to be sufficiently small compared to Planck scale, which is easily
guaranteed from (\ref{condvalue}) for sufficiently large negative
values of the dilaton field $\phi_0$ that characterizes inflation in the scenario of ~\cite{emnstaro}.
Given that in our string scenarios 
the bulk cosmological constant ${\tilde \Lambda} $ is negative~\cite{emnstaro}, (\ref{neglambda}), 
it is possible that the bulk
density of D-particles during the de Sitter phase is such that in order of magnitude
 \begin{equation}
 \mathcal{G} \gg -1 \qquad, \quad {\rm e.g} \quad \frac{T_3}{g_{s0}}\,
 e^{\phi_0} \sim \frac{2\, |{\tilde \Lambda} |}{3\, \kappa_0}~,
 \end{equation}
in which case Eq.~(\ref{condvalue}) implies:
\begin{equation}\label{condvalue2}
{\tilde \alpha} \simeq 2\, e^{2\phi_0} \, \frac{T_3}{g_{s0}}\,
e^{\phi_0} \sim e^{2\phi_0 }\, \frac{4\,|{\tilde \Lambda} |}{3\,
  \kappa_0}~,
\end{equation}
and the condensate ${\tilde \alpha} \ll 1$ (as required for
consistency of the approach) for $|{\tilde \Lambda} |/\kappa_0 \ll 1$
and \emph{finite}, but large negative values of $\phi_0$~\footnote{We remind the reader that we 
are interested in the slow-roll phase of the dilaton, in which $\phi_0$ is almost constant over 
appropriate time scales considered here.}. 

In such a case, the formation of the condensate may be understood as
stabilizing the brane vacuum. Indeed, as follows from
(\ref{action1}), upon the formation of constant quantum vacuum
condensates ${\tilde \alpha}$, the dark energy terms in the brane
effective action (in the Einstein frame) assume the form
\begin{equation}
{\rm D3-Brane~Dark~Energy} \, \sim \, {\tilde \alpha} + e^{2\phi_0}
\left(\frac{T_3}{g_{s0}}\, e^{\phi_0} - \frac{|{\tilde \Lambda}
  |}{\kappa_0} \right) ~.
\label{de}
\end{equation}
 In the absence of a condensate, the brane vacuum energy would be
 \emph{negative}, of order $-|{\tilde \Lambda} |/\kappa_0$, which
 would indicate an instability of the vacuum. The true stable vacuum
 of the theory would then be the one in which the condensate forms,
 with the value (\ref{condvalue2}), which implies a
 positive (de~Sitter-type) \emph{vacuum energy} (\ref{de}) of order 
 \bea\label{condensenergy}
%{\rm D3-Vacuum-Energy} \sim  
  \frac{e^{2\phi_0}\,|{\tilde \Lambda}
 |}{\kappa_0 } \, > \, 0, 
\eea 
which is small for large negative values of the dilaton $\phi_0 << 0$. Negative values of the condensate
can also be found in a way consistent with the stabilization of the vacuum,
in the sense that the total vacuum energy (\ref{de}) is positive, 
and lower than the value when the condensate was absent. 
In this case, the parameters $T_3$ and $\tilde \Lambda$ assume values such that the total 
vacuum energy on the brane is positive.

\section{D-particle-Vacuum condensates and perturbations during Dilaton inflation \label{sec:cone}}

We now discuss the possible effects of quantum fluctuations of the condensate field $\tilde a(t)$ 
on the curvature as well as on the tensor perturbations during 
the dilaton-driven de Sitter phase. 
As mentioned above, one may assume, without loss of generality, that
in this case the dilaton assumes some large, negative and almost constant value, $\phi_0$, 
which characterizes its slow-roll phase. Hence, in the following it suffices qualitatively to 
 concentrate only on the effective action obtained from 
(\ref{action1}) by truncating the Born-Infeld expansion 
to second order in the recoil vector field $A_\mu$ and replacing $F_{\mu\nu}F^{\mu\nu}$ 
by the condensate quantum fluctuating field  $\langle \frac{1}{4} F_{\mu\nu} F^{\mu\nu} \rangle \sim \tilde \alpha (t)$. 

According to the discussion in \cite{emnstaro}, we can also add to the effective action
in the Einstein frame (w.r.t. to the dilaton factors) a 
dilaton-independent positive contribution ${\mathcal A} > 0$ to the brane vacuum energy due to classical
effects of ensembles of D-particles, which, according to the discussion in \cite{emnstaro}, is essential for ensuring 
a de Sitter phase driven by the dilaton. The quantity ${\mathcal A}$ is proportional to the 
density of D-particles on the brane during the inflationary period. 

In this work, we start with the effective action in the string frame~\cite{yusaf}, 
and then pass to the Einstein frame. The reason, as we shall see, is that in our case the passage to the 
Einstein frame requires a conformal rescaling by a combination of scalar fields, 
the dilaton $\phi_0$ and a condensate emerging from the recoil degrees of freedom of the D-particle
ensemble during their interaction with the string matter. In the string frame, 
the vacuum energy due to the D-particle ensemble depends on the dilaton as 
${\mathcal A}_s \equiv \frac{M_s}{g_{s0}} e^{-\phi_0} \, n_s$,
with $n_s$ the proper space density of D-particles in the string frame.
In the Einstein frame (w.r.t. dilaton factors) ${\mathcal A}_s$ yields the aforementioned 
dilaton-independent cosmological constant ${\mathcal A}$. 
Thus, the string-frame effective action we consider, 
in the dilaton-induced inflationary period with near-constant $\phi_0$, is~\cite{yusaf}:
\bea\label{effac}
&& S^{(4)}_{\rm eff} \simeq \nonumber \\ 
&& \int d^4 x \,\sqrt{-g}\, \Big[
  -\Big(\frac{T_3}{g_{s0}}\,e^{-\phi_0} + e^{-2\phi_0} \,\frac{1}{\kappa_0}
  \,{\tilde \Lambda} + \mathcal{A}_s \Big) - \frac{1}{4}{\tilde \alpha}(x) + \nonumber \\
 &&  \left(\alpha \frac{T_3}{g_{s0}}\,
  e^{-\phi_0} + \frac{1}{\kappa_0} \, e^{-2\phi_0} + 
  \frac{\alpha}{4} \, {\tilde \alpha}(x)\right)\, R(g)  \Big]\, + \dots \equiv \nonumber \\
 && \int d^4 x\, \sqrt{-g} \,\frac{e^{-2\phi_0}}{2\kappa^2_{\rm eff}} \Big[ \Big(1 + 2\sigma(x) \Big) R - 2\Lambda \sigma(x) - 2\tilde{\mathcal{B}} \Big] + \dots, \nonumber \\ 
\eea
where $2\kappa^2_{\rm eff} \equiv  \Big(\alpha \frac{T_3 e^{\phi_0}}{g_{s0}} + 
\frac{1}{\kappa_0}\Big)^{-1} \simeq \kappa_0$ for the large negative values of $\phi_0$ 
characterizing the inflationary regime~\cite{emnstaro}, 
is the four-dimensional reduced Planck constant, $\kappa_{\rm eff}^2 = M_{\rm Pl}^{-2}$,
to be used in phenomenological studies,
$0 < \Lambda \equiv \frac{\kappa_{\rm eff}^2}{\alpha} \propto M_s^2/M_{\rm Pl}^2$
is a positive constant to be determined by comparison with data, 
\begin{equation}\label{cosmoconst}
2{\tilde {\mathcal B}} \equiv \frac{T_3}{g_{s0}}\,e^{\phi_0} - \frac{1}{\kappa_0}
  \,|{\tilde \Lambda}| + e^{2\phi_0}\,\mathcal{A}_s~,
\end{equation}  
is an effective cosmological constant, 
and 
\bea\label{condfield}
\sigma(x) \equiv \frac{1}{4} \alpha \kappa_{\rm eff}^2 \, e^{2\phi_0}\, {\tilde \alpha}(x)
\eea 
is a dimensionless condensate field.
We remind the reader that the parameter $\alpha \propto \alpha^\prime > 0$, e.g., in the example of~\cite{Cheung}
$\alpha = \frac{\pi^2}{6} \alpha^\prime$.

The action (\ref{effac}) describes the dynamics of two scalar fields interacting with gravity,
namely the slowly-rolling dilaton $\phi_0$ and the condensate $\sigma$.
%In this sense to study inflation one may follow the 
%approach of \cite{WZ}, by also evaluating properly the non-Gaussianity parameters $f_{NL}$.
%We postpone such a complete analysis for a future publication. 
In what follows we restrict ourselves to the effects of the fluctuations of the D-particle-induced 
condensate field $\sigma(x)$ on the perturbations during the inflationary phase induced by the dilaton,
considering the dilaton field in (\ref{effac}) as approximately constant.  
We shall work in units of the reduced Planck constant, i.e., we set $\kappa^2_{\rm eff} \simeq \kappa_0 =1$. 

We pass into the Einstein frame, denoted by a supersctipt $E$, by redefining the metric~\cite{aben}:
\begin{eqnarray}\label{confmetric}
g_{\mu\nu} \rightarrow g^E_{\mu\nu} & = & \left(1 + 2 {\sigma (x)} \right) \, e^{-2\phi_0} \, g_{\mu\nu} ~, 
\end{eqnarray}
in which case the field $\sigma(x)$ becomes a dynamical scalar degree of freedom. 
We define a canonically-normalized scalar field $\varphi(x)$ 
\bea
  \varphi (x) & \equiv & \sqrt{\frac{3}{2}} \, {\rm ln} \, \left(1 + 2\, {\sigma \left(x\right)} \right)~,
\eea
so that the action (\ref{effac}) becomes
\bea\label{steps}
	S^{(4}_{\rm eff} &=&  \frac{1}{2}\,\int d^4 x \sqrt{-g^E}\,  \left(R^E +  g^{E\, \mu\, \nu} \, \partial_\mu \, \varphi \, \partial_\nu \, \varphi - V\right(\varphi\left) \right) + \nonumber \\
&& {\mathcal O}\Big(\partial_\mu \varphi \partial_\nu \phi_0 g^{E \mu\nu}, \partial_\mu \phi_0 \partial_\nu \phi_0 g^{E \mu\nu} \Big)~,	
\eea
with the effective potential $V(\varphi)$ in the inflationary regime of \emph{large negative values} of $\phi_0$~\cite{emnstaro}  given approximately by:
\begin{eqnarray}\label{staropotent}
 V(\varphi ) \simeq \Big(e^{\sqrt{\frac{2}{3}}\varphi} - 1 \Big)  \, \Lambda e^{-2\phi_0}  + 2{\mathcal B} e^{2\phi_0 - 2\sqrt{\frac{2}{3}}\varphi}~, 
 \end{eqnarray}
 where 
\bea
2{\mathcal B} \simeq   - \frac{1}{\kappa_0}
  \,|{\tilde \Lambda}| + e^{-2\phi_0}\,\mathcal{A}~.
 \eea
It is important to note that, in order to arrive at (\ref{staropotent}),
we took into account the conformal nature of the condensate
${\tilde \alpha}(x) = \frac{1}{4} \langle F_{\mu\nu} F_{\rho\sigma} \rangle g^{\mu\rho} g^{\nu\sigma} $
and have ignored terms that are more than quadratic in the vector potential. 
Moreover, as already emphasized, for our purposes here we concentrate on the slow-roll phase of the 
dilaton field $\phi_0$, so any potential-like terms with dilaton time-derivative factors will be ignored. 
In this approximation we need not worry about the cross-kinetic-terms $\partial_\mu \phi_0 \partial^\mu \varphi$, 
which can in any case be eliminated by a further redefinition (mixing) of the fields $\varphi$ and $\phi_0$. 

There is one more ingredient in the effective action (\ref{steps})
that is important for inflationary phenomenology, as we now discuss. This arises
because in brane worlds there are other flux gauge fields, in addition to the recoil vector field, 
that can also be described by appropriate Born-Infeld actions on the D3-brane Universe, and may 
condense~\cite{odintsov} during the dilaton-induced inflationary period. Such condensates 
are also conformally invariant, just like the recoil vector fields above, leading to additional 
vacuum energy terms in the four-dimensional effective action of the following form in the Einstein frame:
$\int d^4 x \sqrt{-g^E} \frac{1}{4} \langle {\mathcal G}_{\mu\nu} {\mathcal G}^{\mu\nu} \rangle^E  = \int d^4 x \sqrt{-g^E } \, {\mathcal D}$,
where ${\mathcal D}$ is a constant, independent of the D-particle recoil condensate field and its fluctuations $\varphi(x)$.
This would therefore lead to an extra (positive) cosmological constant contribution
to the effective potential:
\begin{eqnarray}\label{staropotent2}
&& V(\varphi ) = \Big(e^{\sqrt{\frac{2}{3}}\varphi} - 1 \Big) \, e^{-2\phi_0} \, \Lambda -\frac{|{\tilde \Lambda}|}{\kappa_0} \, e^{2\phi_0 -2\sqrt{\frac{2}{3}}\varphi} + \nonumber \\
&& {\mathcal A}\, e^{-2\sqrt{\frac{2}{3}}\varphi}  + {\mathcal D}~.
 \end{eqnarray}
From the discussion in \cite{emnstaro}, an extra factor $\sqrt{2}$ needs to be absorbed
into the dilaton normalization in order to obtain a canonical kinetic term, yielding finally 
 \begin{eqnarray}\label{staropotent3}
&& V(\varphi ) = \Big(e^{\sqrt{\frac{2}{3}}\varphi} - 1 \Big) \, e^{-\sqrt{2}\phi_0} \, \Lambda -\frac{|{\tilde \Lambda}|}{\kappa_0} \, e^{\sqrt{2}\phi_0 -2\sqrt{\frac{2}{3}}\varphi} + \nonumber \\
&& {\mathcal A}\, e^{-2\sqrt{\frac{2}{3}}\varphi}  + {\mathcal D}~.
 \end{eqnarray} 
For weak condensates $\varphi \ll 1$, where the approximations in this article hold, 
and large negative values of the dilaton $\phi_0 \ll 0$, as appropriate for the slow-roll inflation of \cite{emnstaro}, 
the reader will recognize in (\ref{staropotent3}) the Starobinsky-like
form (\ref{phenostar}) of the effective action 
for the dilaton-driven inflation of ref.~\cite{emnstaro}, provided ${\mathcal A} + {\mathcal D} > 0$.
This can fit the Planck data~\cite{Planck} - though not the BICEP2 data~\cite{Bicep2}
- due to the very small value of the tensor-to-scalar ratio $r$ predicted by this class of theories. 

\begin{figure}[h!!!]
\centering
\includegraphics[width=0.5\textwidth]{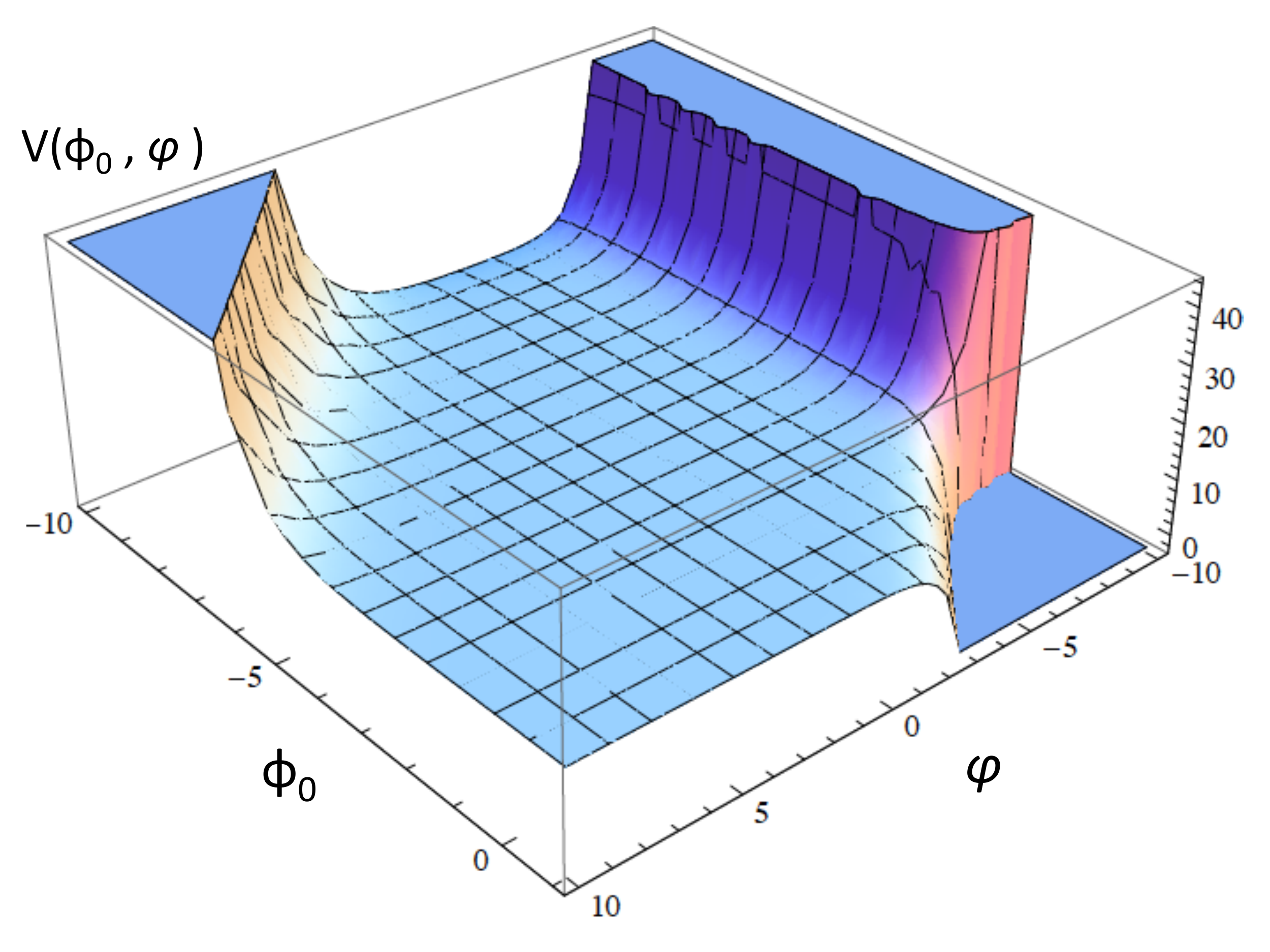} 
		%\includegraphics[width=0.6\textwidth]{Potential.pdf} 
%		\vfill
%\includegraphics[width=0.3\textwidth]{epsilon.pdf} \hfill \includegraphics[width=0.3\textwidth]{epsilon2.pdf}		
%		\vfill		
%		\includegraphics[width=0.3\textwidth]{ns.pdf}				
		\caption{\it The effective  potential (\ref{staropotent2}) as a function of the collective scalar field $\varphi$ representing quantum fluctuations of D-particle condensates and  
		the dilaton $\phi_0$ 
		during its slow-roll inflationary phase, as obtained by choosing
		some representative values, $\Lambda = 10^{-6}$,  
		$\frac{|\tilde{\Lambda}|}{\kappa_0} = 10^{-2},$  ${\mathcal A} = 10^{-4} $ and $\mathcal{D}= 15$ in units of an overall unspecified scale.} \label{fig:potstar2}	
\end{figure}

The form of the effective potential as a function of both the dilaton $\phi_0$ and the
condensate fields field $\varphi$ is illustrated in Fig.~\ref{fig:potstar2},
where we see that a weak condensate: $\varphi \ll 1$ is consistent with the minimization
at large negative $\phi_0$, which is a consistency check of our approach. 
It should be understood that a full analysis of the two-field effective action (\ref{effac}), 
together with a study of non-Gaussianities, along the lines of \cite{WZ}, is required before
a conclusive comparison with data can be made. However, it is clear from the above analysis that 
the effects of D-particle recoil through the formation of appropriate condensates, 
although compatible with an inflationary phase driven by the dilaton,
cannot lead to significant enhancement of the tensor to scalar ratio
if the scalar degrees of freedom associated with the condensate and the dilaton are treated as independent fields. 

On the other hand, we expect the condensate field $\varphi$ to mix with the dilaton via kinetic terms,
so that
\begin{equation}
\label{screen}
\varphi = {\cal C} \sqrt{2} \phi_0 (t) + \alpha (t) 
\end{equation}
where ${\cal C}$ is some undetermined constant and $\alpha (t)$ is independent of the dilaton.
Looking at the exponents in the action (\ref{staropotent3}), we observe that the dilaton effects 
in the second term of the effective potential (\ref{staropotent3}) may then be screened, 
so that the coefficient of $\phi_0$ in the exponential factor becomes significantly smaller than $\sqrt{2}$
for a suitable value of ${\cal C}= {\mathcal O}(1) > 0$.
Specifically, one may obtain an enhanced tensor-to-scalar ratio $r$
that is more compatible with BICEP2~\cite{Bicep2}, if one assumes suitable
values of the model parameters ${\mathcal A}$ and $\Lambda$ 
in (\ref{staropotent3}) so that the inflationary potential for $\phi_0$ values of
interest is dominated by an effective constant that is approached exponentially
with a small coefficient $B$ in the notation of (\ref{phenostar}).
According to (\ref{crucial}), this would then lead to an enhanced value of $r$, and a BICEP2-friendly
value is possible if ${\cal C} \simeq 0.55$.

\section{Hawking Radiation Effects on Tensor Perturbations \label{sec:hawking}}

In this section we mention another way in which tensor perturbations may be enhanced
in our D-particle foam scenario~\cite{emnstaro}. At the end of the dilaton-induced inflation,
this scenario yields a new `out' vacuum state, with a significant population of D-particles. 
Such populations also existed during the inflationary phase, and were responsible for
inducing a non-zero cosmological constant in the string effective action $A$ (\ref{phenostar}).
Moreover, it has been demonstrated~\cite{sarkarnem} that when the D-particle vacuum 
is viewed as an `out' vacuum in an expanding universe, there is particle production,
and the D-particle vacuum,
with stochastic recoil-velocity fluctuations, constitutes a system with an equation of state similar to that of a de Sitter 
space-time (see also~\cite{emnstaro}). 

This particle production has a thermal spectrum \`a la Gibbons-Hawking,
since it is known via non-perturbative string theory~\cite{pbranes} (using duality to a matrix model)
that the thermally excited collection of D0-branes in a D3-brane space-time
constitute a stringy version of a black hole. 
Using this description, the ``scrambling'' time $t_\star$ for this black hole (i.e., the time  
needed for thermalization of the system of ${\mathcal N}$ D0-particles, 
as seen by an observer who is asymptotically far from the horizon of the black hole)
was estimated in~\cite{susskind} in the large ${\mathcal N}$ limit to be:
\be\label{scrambling}
t_\star \sim  \frac{1}{2\pi} T^{-1} \sqrt{-g_{00}(r)} \, {\rm log} (\frac{R_S}{\ell_s})~,
\ee 
where $R_S$ is the Schwarzschild radius of the black hole, and $g_{00}$ is the temporal component of the 
metric:
\be\label{bh}
ds^2 = - \Big(1 - R_S r^{-1} \Big) dt^2 + (\Big(1 - r^{-1} R_S \Big)^{-1} dr^2 + r^2 d\Omega^2 ~.
\ee
where $R_S=2M/M_P^2$ is the 
Schwarschild radius  (with $M_P$ the four-dimensional Planck mass).

In (\ref{scrambling}), $T \propto {\mathcal N}^{-1/2}$ becomes the
Hawking evaporation temperature $T_H = \frac{M_P^2}{8\pi M} = \frac{1}{4\pi R_s} $
for an observer at a large distance $r \to \infty$ from the horizon,
and  the ${\rm log}(R/\ell_s)$ factor is related to  ${\rm log} S$, where 
$S \propto {\rm log}{\mathcal N}$ is the entropy of the black hole. This logarithmic behaviour
of the scrambling time is associated in~\cite{susskind} with the fact that in the
microscopic description of the black hole the relevant degrees of freedom are delocalized as in the matrix model. 

If we assume that D0-branes on the D3-brane Universe thermalize,
a cosmological brane observer would find him/her self inside the horizon of the black hole formed by the 
thermal ensemble of D0-particles~\cite{susskind}. 
There is a duality between a de Sitter space-time with radius $R=1/H$ 
and Gibbons-Hawking temperature $T_{GH} = H/2\pi$, 
where $H$ is the Hubble parameter, parametrized by the metric 
\be\label{desitter2}
ds^2 = - \Big(1 - r^2 H^2 \Big) dt^2 + (\Big(1 - r^2 H^2 \Big)^{-1} dr^2 + r^2 d\Omega^2 ~.
\ee
and a Schwarzschild Black hole of mass $M$ with metric (\ref{bh}).

This duality extends the scrambling time~\cite{susskind}, 
with the Hawking temperature of the black hole corresponding to the Gibbons-Hawking
temperature. 
The scrambling time of de Sitter space-time has also been estimated using matrix theory in~\cite{susskind},
and found to be analogous to that of the black hole (\ref{scrambling}):
\be\label{dsscramble}
t_\star^{{\rm dS}} \sim \hbar T^{-1} _{GH} \, {\rm log}(1/(H\ell_s)) \propto \frac{1}{H} {\rm log} (1/(H\ell_s)) \, ,
\ee
where $\ell_s$ is the string length. 

If the Hubble parameter during inflation is of order $10^{-5} M_P$,
the maximum allowed by the BICEP2 data (\ref{hubblebicep2}),  
and assuming that the string length is of order the four-dimensional Planck length, $\ell_s \sim \ell_P = M_P^{-1}$, 
the thermalization time is of order $t_\star \sim  10^5 \, {\rm log}(10^5 \, \frac{M_s}{M_P}) M_P^{-1}$,
which is plausibly less than the duration of inflation in realistic theories, $\sim 10^9 M_P^{-1}$. 
Thus, the D-particles and the out vacuum are thermalized at the end of the de Sitter phase.

It was suggested in~\cite{mohanty} that, at the exit from a generic inflationary de-Sitter vacuum,
the vacuum structure changes to a new `out' vacuum in which there is mixing between the de Sitter
modes and the eigenmodes of the `out' vacuum. Assuming that the Bogolyubov coefficients of this mixing
have a thermal distribution with temperature $T=H_\star /2\pi$, it was shown in~\cite{mohanty} that 
the two-point quantum correlation functions of (transverse, traceless)  graviton tensor modes in the `out' vacuum state
would have the spectrum
\be\label{tspec}
P_T \simeq \frac{2\, H^2_\star}{\pi^2 \, M_{\rm Pl}^2}  \Big(\frac{k}{a \, H_\star}\Big)^{-2\epsilon} 
\Big(\frac{\pi}{2} \, \frac{k}{a H_\star}\Big) 
\ee
for the scales $k \ll a H_\star$ of relevance to observations, providing a contribution to
the tensor perturbation spectrum with spectral index
\bea\label{nt}
n_T = dP_T/d{\rm ln}(k) \sim 1 - 2\epsilon~.
\eea
On the other hand, dilaton-induced inflation proceeds in such a way that the 
$\eta$ parameter is negative, corresponding to tachyonic scalar curvature perturbations.
These are not thermalized, so the scalar perturbation spectrum is not affected, and assumes the standard form.
Hence the scalar-to-tensor ratio for this case has a scale-dependent form that may be different from the standard
result. 

\section{Conclusions \label{sec:concl}}

In this article we have explored how D0-brane defects in the stringy cosmological scenario
of~\cite{emnstaro} may affect tensor perturbations so as to yield a larger tensor-to-scalar ratio.

One possibility is the formation of condensates of the vector field representing quantum fluctuations in
the recoil velocities of D0-branes on the D3-brane Universe (or some other fields with a Born-Infeld action). 
These condensates may contain terms proportional to the dilaton field,
due to the presence of mixed kinetic terms in the effective lagrangian,
which in turn leads to effective potentials of the type (\ref{phenostar}) 
but with exponents $B$ much smaller than the Starobinsky value. 
According then to (\ref{crucial}), an enhanced ratio $r$ of tensor to scalar perturbations may be obtained.

Another possibility is that D-particle defects may generate a non-trivial `out' vacuum,
characterized by thermal spectrum particle production, with modes that mix with the de Sitter modes
to enhance the tensor-to-scalar ratio and modify the spectrum of tensor perturbations.
The blue-tilted spectrum $n_T \sim 1$ of tensor perturbations suggested in this second scenario
may be in tension with other constraints,
as pointed out in \cite{bbnconstr}, but we leave for future work their analysis in the context of a detailed brane cosmology. 

We close by remarking that, although the above ideas are certainly speculative,
they may provide sufficient motivation to study such stringy-inflation
scenarios more deeply, with a view to accommodating current astrophysical data as
well as Planck and BICEP2 data.

%%%%%%%%%%%%%%%%%%%%%%%%%%%%%%%%%%%%%%%%%

\section*{Acknowledgements}

The work  of J.E. and 
N.E.M. was supported in part by the London Centre for Terauniverse Studies (LCTS), using funding from the European Research Council via the Advanced Investigator Grant 267352 and by STFC (UK) under the research grant ST/J002798/1.
That of D.V.N.  was supported in part by the DOE grant DE-FG03-95-ER-40917.


\begin{thebibliography}{99}

%\cite{Ellis:2014cma}
\bibitem{emnstaro} 
J.~Ellis, N.~E.~Mavromatos and D.~V.~Nanopoulos,
  %``Starobinsky-Like Inflation in Dilaton-Brane Cosmology,''
  Phys.\ Lett.\ B {\bf 732}, 380 (2014)
  [arXiv:1402.5075 [hep-th]].
  %%CITATION = ARXIV:1402.5075;%%
  %5 citations counted in INSPIRE as of 27 May 2014

%\cite{Starobinsky:1980te}
\bibitem{staro} 
  A.~A.~Starobinsky,
  %``A New Type of Isotropic Cosmological Models Without Singularity,''
  Phys.\ Lett.\ B {\bf 91}, 99 (1980).
  %%CITATION = PHLTA,B91,99;%%
  %1685 citations counted in INSPIRE as of 14 Nov 2013
  
 \bibitem{BS} 
  F.~L.~Bezrukov and M.~Shaposhnikov,
  %``The Standard Model Higgs boson as the inflaton,''
  Phys.\ Lett.\ B {\bf 659} (2008) 703
  [arXiv:0710.3755 [hep-th]].
  %%CITATION = ARXIV:0710.3755;%%
  %435 citations counted in INSPIRE as of 02 Jun 2014

  \bibitem{olive1} J.~Ellis, D.~V.~Nanopoulos and K.~A.~Olive,
  %``No-Scale Supergravity Realization of the Starobinsky Model of Inflation,''
  Phys.\ Rev.\ Lett.\  {\bf 111} (2013) 111301
   [Erratum-ibid.\  {\bf 111} (2013) 12,  129902]
  [arXiv:1305.1247 [hep-th]].
  %%CITATION = ARXIV:1305.1247;%%
  %72 citations counted in INSPIRE as of 02 Jun 2014
  
  \bibitem{KL}
  R.~Kallosh and A.~Linde,
  %``Superconformal generalizations of the Starobinsky model,''
  JCAP {\bf 1306} (2013) 028
  [arXiv:1306.3214 [hep-th]].
  %%CITATION = ARXIV:1306.3214;%%
  %68 citations counted in INSPIRE as of 02 Jun 2014
  
  \bibitem{olive2} J.~Ellis, D.~V.~Nanopoulos and K.~A.~Olive,
%``Starobinsky-like Inflationary Models as Avatars of No-Scale Supergravity,''
  JCAP {\bf 1310}, 009 (2013)
  [arXiv:1307.3537 [hep-th]].
  %%CITATION = ARXIV:1307.3537;%%
  %16 citations counted in INSPIRE as of 14 Nov 2013


  
\bibitem{Planck} 
P.~A.~R.~Ade {\it et al.}  [Planck Collaboration],
  %``Planck 2013 results. XXII. Constraints on inflation,''
  arXiv:1303.5082 [astro-ph.CO]
  %%CITATION = ARXIV:1303.5082;%%
 and
  %``Planck 2013 results. I. Overview of products and scientific results,''
  arXiv:1303.5062 [astro-ph.CO].
  %%CITATION = ARXIV:1303.5062;%%
  

  
  %\cite{Ade:2014xna}
\bibitem{Bicep2} 
  P.~A.~R.~Ade {\it et al.}  [BICEP2 Collaboration],
  %``BICEP2 I: Detection Of B-mode Polarization at Degree Angular Scales,''
  arXiv:1403.3985 [astro-ph.CO].
  %%CITATION = ARXIV:1403.3985;%%
  %102 citations counted in INSPIRE as of 02 Apr 2014
  
  
  
  
  
\bibitem{encyclo}  J.~Martin, C.~Ringeval and V.~Vennin,
  %``Encyclopaedia Inflationaris,''
  arXiv:1303.3787 [astro-ph.CO].
  %%CITATION = ARXIV:1303.3787;%%
  %65 citations counted in INSPIRE as of 13 Apr 2014

\bibitem{Subir}
  H.~Liu, P.~Mertsch and S.~Sarkar,
  %``Fingerprints of Galactic Loop I on the Cosmic Microwave Background,''
  arXiv:1404.1899 [astro-ph.CO].
  %%CITATION = ARXIV:1404.1899;%%
  %9 citations counted in INSPIRE as of 27 May 2014
  
\bibitem{MS}
M.~J.~Mortonson and U.~Seljak,
  %``A joint analysis of Planck and BICEP2 B modes including dust polarization uncertainty,''
  arXiv:1405.5857 [astro-ph.CO].
  %%CITATION = ARXIV:1405.5857;%%
  %2 citations counted in INSPIRE as of 27 May 2014
  
%\cite{Flauger:2014qra}
\bibitem{Flauger} 
  R.~Flauger, J.~C.~Hill and D.~N.~Spergel,
  %``Toward an Understanding of Foreground Emission in the BICEP2 Region,''
  arXiv:1405.7351 [astro-ph.CO].
  %%CITATION = ARXIV:1405.7351;%%
  
 %\cite{Mavromatos:2012ha}
\bibitem{yusaf} 
  N.~E.~Mavromatos, M.~Sakellariadou and M.~F.~Yusaf,
  %``Stringy Models of Modified Gravity: Space-time defects and Structure Formation,''
  JCAP {\bf 1303}, 015 (2013)
  [arXiv:1211.1726 [hep-th]].
  %%CITATION = ARXIV:1211.1726;%%
  %1 citations counted in INSPIRE as of 04 Apr 2014
  
  
  \bibitem{Cheung}
  Y.~K.~Cheung, M.~Laidlaw and K.~Savvidy,
  %``Open string gravity?,''
  JHEP {\bf 0412}, 028 (2004)
  [arXiv:hep-th/0406245].
  
  
  \bibitem{recoil}  I.~I.~Kogan and N.~E.~Mavromatos,
  %``World sheet logarithmic operators and target space symmetries in string theory,''
  Phys.\ Lett.\ B {\bf 375}, 111 (1996)
  [hep-th/9512210].
  %%CITATION = HEP-TH/9512210;%%
  %107 citations counted in INSPIRE as of 11 Feb 2014 
   I.~I.~Kogan, N.~E.~Mavromatos and J.~F.~Wheater,
  %``D-brane recoil and logarithmic operators,''
  Phys.\ Lett.\ B {\bf 387}, 483 (1996)
  [hep-th/9606102].
  %%CITATION = HEP-TH/9606102;%%
  %127 citations counted in INSPIRE as of 11 Feb 2014
  J.~R.~Ellis, N.~E.~Mavromatos and D.~V.~Nanopoulos,
  %``D-brane recoil mislays information,''
  Int.\ J.\ Mod.\ Phys.\ A {\bf 13}, 1059 (1998)
  [hep-th/9609238].
  %%CITATION = HEP-TH/9609238;%%
  %70 citations counted in INSPIRE as of 11 Feb 2014
  
  \bibitem{foam}
  J.~R.~Ellis, N.~E.~Mavromatos and M.~Westmuckett,
  %``A supersymmetric D-brane model of space-time foam,''
  Phys.\ Rev.\  D {\bf 70} (2004) 044036
  [arXiv:gr-qc/0405066];
   J.~R.~Ellis, N.~E.~Mavromatos, D.~V.~Nanopoulos and M.~Westmuckett,
  %``Liouville cosmology at zero and finite temperatures,''
  Int.\ J.\ Mod.\ Phys.\  A {\bf 21} (2006) 1379
  [arXiv:gr-qc/0508105];
  

\bibitem{foam2} 
  N.~E.~Mavromatos, V.~A.~Mitsou, S.~Sarkar and A.~Vergou,
  %``Implications of a Stochastic Microscopic Finsler Cosmology,''
  Eur.\ Phys.\ J.\ C {\bf 72}, 1956 (2012)
  [arXiv:1012.4094 [hep-ph]].
  %%CITATION = ARXIV:1012.4094;%%
  
  
  
  
  \bibitem{odintsov} 
  E.~Elizalde, J.~E.~Lidsey, S.~Nojiri and S.~D.~Odintsov,
  %``Dirac-Born-Infeld quantum condensate as dark energy in the universe,''
  Phys.\ Lett.\ B {\bf 574}, 1 (2003)
  [hep-th/0307177].
  %%CITATION = HEP-TH/0307177;%%
  
  \bibitem{aben} 
  I.~Antoniadis, C.~Bachas, J.~R.~Ellis and D.~V.~Nanopoulos,
  %``Comments on cosmological string solutions,''
  Phys.\ Lett.\ B {\bf 257}, 278 (1991);
  %%CITATION = PHLTA,B257,278;%%
  %95 citations counted in INSPIRE as of 09 Feb 2014
 %``An Expanding Universe in String Theory,''
  Nucl.\ Phys.\ B {\bf 328}, 117 (1989);
  %%CITATION = NUPHA,B328,117;%%
  %261 citations counted in INSPIRE as of 09 Feb 2014
%``Cosmological String Theories and Discrete Inflation,''
  Phys.\ Lett.\ B {\bf 211}, 393 (1988).
  %%CITATION = PHLTA,B211,393;%%
  %266 citations counted in INSPIRE as of 09 Feb 2014
  
   %\cite{Ellis:2014rja}
\bibitem{WZ} 
    D.~Croon, J.~Ellis and N.~E.~Mavromatos,
  %``Wess-Zumino Inflation in Light of Planck,''
  Phys.\ Lett.\ B {\bf 724}, , 165 (2013)
  [arXiv:1303.6253 [astro-ph.CO]];
  %%CITATION = ARXIV:1303.6253;%%
  %17 citations counted in INSPIRE as of 09 Feb 2014
  J.~Ellis, N.~E.~Mavromatos and D.~J.~Mulryne,
  %``Exploring Two-Field Inflation in the Wess-Zumino Model,''
  arXiv:1401.6078 [astro-ph.CO].
  %%CITATION = ARXIV:1401.6078;%%


 
   %\cite{Mavromatos:2007ak}
\bibitem{sarkarnem} 
N.~E.~Mavromatos and J.~Papavassiliou,
  %``Superheavy dark matter anisotropies from D particles in the early universe,''
  Int.\ J.\ Mod.\ Phys.\ A {\bf 19}, 2355 (2004)
  [hep-th/0307028];
  %%CITATION = HEP-TH/0307028;%%
  %4 citations counted in INSPIRE as of 02 Jun 2014
N.~E.~Mavromatos and Sarben~Sarkar,
  %``Towards a microscopic construction of flavour vacua from a space-time foam model,''
  New J.\ Phys.\  {\bf 10}, 073009 (2008)
  [arXiv:0710.4541 [hep-th]].
  %%CITATION = ARXIV:0710.4541;%%
  %20 citations counted in INSPIRE as of 13 Apr 2014

%\cite{Horowitz:1991cd}
\bibitem{pbranes} 
  G.~T.~Horowitz and A.~Strominger,
  %``Black strings and P-branes,''
  Nucl.\ Phys.\ B {\bf 360}, 197 (1991).
  %%CITATION = NUPHA,B360,197;%%
  %845 citations counted in INSPIRE as of 14 Apr 2014

  
   %\cite{Sekino:2008he}
\bibitem{susskind} 
 Y.~Sekino and L.~Susskind,
  %``Fast Scramblers,''
  JHEP {\bf 0810}, 065 (2008)
  [arXiv:0808.2096 [hep-th]].
  %%CITATION = ARXIV:0808.2096;%%
  %106 citations counted in INSPIRE as of 14 Apr 2014
  L.~Susskind,
  %``Addendum to Fast Scramblers,''
  arXiv:1101.6048 [hep-th].
  %%CITATION = ARXIV:1101.6048;%%
  %28 citations counted in INSPIRE as of 14 Apr 2014
  
  
   %\cite{Mohanty:2014kwa}
\bibitem{mohanty}   
S.~Mohanty and A.~Nautiyal,
  %``Signature of Gibbons-Hawking temperature in the BICEP2 measurement of gravitational waves,''
  arXiv:1404.2222 [hep-ph].
  %%CITATION = ARXIV:1404.2222;%%
  
  
   \bibitem{bbnconstr}  K.~M.~Smith, C.~Dvorkin, L.~Boyle, N.~Turok, M.~Halpern, G.~Hinshaw and B.~Gold,
  %``On quantifying and resolving the BICEP2/Planck tension over gravitational waves,''
  arXiv:1404.0373 [astro-ph.CO].
  %%CITATION = ARXIV:1404.0373;%%
  %6 citations counted in INSPIRE as of 13 Apr 2014
  
   
  
\end{thebibliography}
  \end{document}